\documentclass[12pt]{article}
\usepackage{epsfig,axodraw}
\begin{document}

\title{\bf
Six-Quark Amplitudes from Fermionic MHV Vertices}

\author{Xun Su\thanks{Supported in part by fund from the National
Natural Science Foundation of China with grant Number 90103004.} \\
Institute of Theoretical Physics,
Chinese Academy of Sciences\\
P. O. Box 2735,  Beijing 100080, P. R. China\\ e-mail: ghkcn@itp.ac.cn \\  \\
Jun-Bao Wu \\ School of Physics, Peking University \\
Beijing 100871, P. R. China\\  e-mail: jbwu@pku.edu.cn}

\maketitle
\begin{abstract}
The fermionic extension of the CSW approach to perturbative gauge
theory coupled with fermions is used to compute  the six-quark QCD
amplitudes. We find complete agreement with the results obtained
by using the usual Feynman rules.
\end{abstract}

\newpage
\section{Introduction}
Recently Witten \cite{Wittenb} found a deep connection between the
perturbative gauge theory and   string theory in twistor space
\cite{Penrosea}. Based on this work, Cachazo, Svrcek and Witten
(CSW for short) reformulated the  perturbative calculation of the
scattering amplitudes in Yang-Mills theory using the off shell MHV
vertices \cite{Wittena}. The MHV vertices are the usual tree level
MHV scattering amplitudes in gauge theory \cite{Parkea, Giele},
continued off shell in a particular fashion as given in
\cite{Wittena}. (For references on perturbative calculations, see
for example \cite{Parkeb}-\cite{Others}. The 2 dimensional origin
of the MHV amplitudes in gauge theory was first given in
\cite{Nair}.) Some sample calculations were done in
\cite{Wittena}, sometimes with the help of symbolic manipulation.
The correctness of the rules was partially verified by reproducing
the known results for small number of gluons \cite{Parkeb}.

In a previous work \cite{WuZhub} (for recent works, see
\cite{Zhu}-\cite{Abe}),   the extension of the CSW approach to
theories with fermions (see also \cite{GeorgiouKhoze}) is
discussed  and used to calculate the googly fermionic amplitudes.
The amplitudes with $3$ quark-anti-quark pairs which are neither
MHV nor googly were also analyzed in that paper.
 In this paper, we will calculate these amplitudes explicitly
by using the CSW rules and show that the results are in agreement
with the results obtained by using the usual Feynman rules
\cite{Gunion}. Although some generic non-MHV fermionic amplitudes
were also calculated in \cite{GeorgiouKhoze,Georgioub}, we found
that it is also worthy to do this calculation. As we will see in
the following, the calculation by using the usual  Feynman rules
is even simpler. So the purpose of our paper is actually to check
that the MHV rules are really correct in this non-trivial case.

The MHV (and googly) amplitudes with gluinos or one
quark-anti-quark pair can be obtained from the gluonic amplitudes
via supersymmetric Ward identities (SWI's)
\cite{Grisarua,Grisarub,Parkeb}. But there are more fermionic
amplitudes which cannot be obtained in this way. In \cite{Parkec},
it has been discussed that neither the non-MHV (googly) amplitudes
with gluinos nor the MHV (googly) amplitudes with two
quark-anti-quark pairs can be obtained  from the gluonic
amplitudes by using the SWI's (See also \cite{Georgioub}). In some
sense the  amplitudes which cannot be obtained via SWI's have more
information. So it is worthy to calculate these amplitudes by
using the CSW rules. Although the CSW rule can be partially
understood from the twistor string theory \cite{Gukov}, a full
understanding of the CSW approach from the conventional field
theory is not reached \cite{Bernc}.

This paper is organized as follows. In section 2, we first review
CSW rules for gauge theory without fermions. Then we review
extended CSW rules for gauge theory with quarks and the analysis
on the CSW diagrams for six-quark amplitudes. In section 3, we
calculate the amplitudes with $3$ quark-anti-quark pairs by using
the fermionic extension of CSW rules. We show that the result
coincides with which from Feynman rules.

\section{CSW rules with fermions}
First let us review the rules for calculating tree level gauge
theory gluonic amplitudes proposed in \cite{Wittena}. Here we
follow the presentation given in \cite{Zhu,WuZhua,WuZhub} closely,
and consider only partial amplitudes \cite{Parkeb}. We will use
the convention that all momenta are outgoing. By MHV (with gluons
only), we always mean an amplitude with precisely two gluons of
negative helicity. If the two gluons of negative helicity are
labelled as $r,s$, the MHV vertices are given as follows:
\begin{equation}
V_n =  {\langle\lambda_r,\lambda_s\rangle^4\over
\prod_{i=1}^n\langle\lambda_i, \lambda_{i+1}\rangle} .
\label{eqone}
\end{equation}
For an on shell (massless) gluon, the momentum in bispinor basis
is given as:
\begin{equation}
p_{a\dot a} = \sigma^\mu_{a\dot a} p_\mu = \lambda_a \tilde{
\lambda}_{\dot a}.
\end{equation}
For an off shell momentum, we can no longer define $\lambda_a$ as
above. The off-shell continuation given in \cite{Wittena} is to
choose an arbitrary spinor $\tilde\eta^{\dot a}$ and then to
define $\lambda_a$ as follows:
\begin{equation}
\lambda_a = p_{a\dot a}\tilde{\eta}^{\dot a}.
\end{equation}
For an on shell momentum $p$, we will use the notation
$\lambda_{pa}$ which is proportional to $\lambda_a$:
\begin{equation}
\lambda_{pa} \equiv  p_{a\dot a} \tilde{\eta}^{\dot a} = \lambda_a
\tilde{\lambda}_{\dot a} \tilde{\eta}^{\dot a} \equiv \lambda_a
\phi_p.
\end{equation}
As demonstrated in \cite{Wittena}, it is consistent to use the
same $\tilde\eta$ for all the off shell momenta. The final result
is independent of $\tilde{\eta}$.

By using only MHV vertices, one can build a tree diagram by
connecting MHV vertices with propagators.  For the propagator of
momentum $p$, we assign a factor $1/p^2$. The helicity at two ends
of a propagator must be opposite. Any possible diagram with the
color factor ${\rm Tr}(T^{a_1}\cdots T^{a_n})$ contributes to the
partial amplitude $A_n(g_1^{h_1},\cdots,g_n^{h_n})$.

Now we review the extension of CSW rules to the gauge theory
coupled with quarks and anti-quarks \cite{GeorgiouKhoze,WuZhub}.
For this theory we can decompose an amplitude into partial
amplitudes with definite color factors \cite{Parkeb}. For
simplicity we will assume that all quarks have different flavors.
When there are identical quarks, the amplitudes can be easily
obtained from the amplitudes with distinct quarks.  Also we will
assume the gauge group to be $U(N)$ instead of $SU(N)$. For a
connected diagram with $m$ pair of quark-antiquark, the color
factor is
\begin{equation}
(T^{a_1}   \cdots T^{a_{n_1}})_{i_1\bar{i_2}} ( T^{a_{n_1+1}}
 \cdots T^{a_{n_2}})_{i_2\bar{i_3}} \cdots (
T^{a_{n_{m-1}+1}}   \cdots T^{a_{n}})_{i_m\bar{i_1}} ,
\end{equation}
for a particular ordering of the quark-antiquarks and gluons
\cite{Mangano}. For amplitudes with connected Feynman diagrams,
the quark-antiquark color indices $(i, \bar i)$ must form a ring
of length exactly $m$. This can be proved by induction with the
number of pairs $m$.

 \begin{figure}[ht]
    \epsfxsize=80mm%
    \hfill\epsfbox{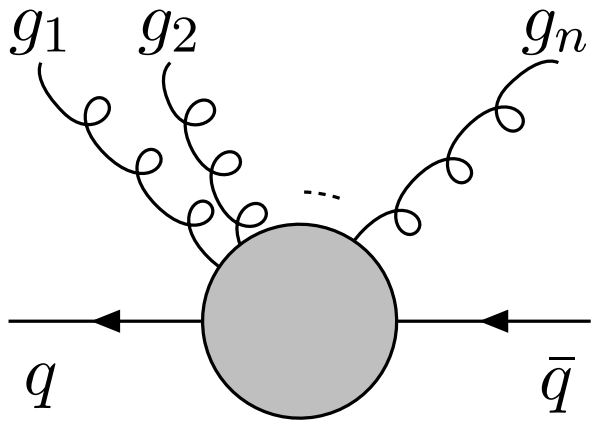}\hfill~\\
    \caption{The graphic representation for the single  pair of
    quark-antiquark partial amplitude. Gluons are emitted from
    one side of the fermion line only.}
     \label{quarkline}
   \end{figure}

For a single quark-antiquark pair the color factor is $ (T^{a_1}
\cdots T^{a_{n}})_{i\bar{i}}$. The partial amplitude is denoted as
$A_{n+2}(\Lambda^h_{q}, g_1, \cdots, g_n, \Lambda^{-h}_{\bar
q})$\footnote{$h$ denotes the helicity of the quark $q$. The
helicity of the antiquark $\bar q$ is $-h$ by helicity
conservation along the quark line.}. It is represented as in
Fig.~\ref{quarkline}. We note that gluon lines are emitted only
from one side of the (connected) quark-anti-quark line. We will
stick to this rule also for multi-pair  of quark-antiquark
diagrams.

 \begin{figure}[ht]
    \epsfxsize=100mm%
    \hfill\epsfbox{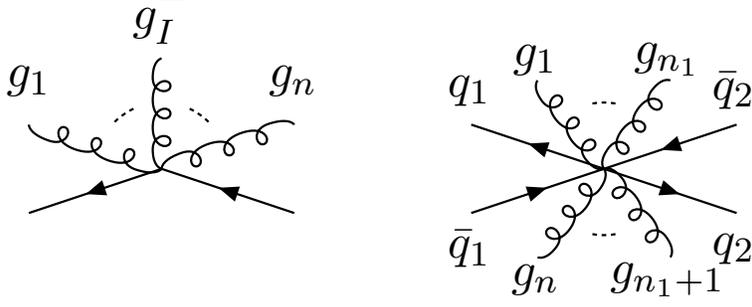}\hfill~\\
    \caption{The 2 MHV vertices with quark-antiquarks.}
         \label{mhvall}
   \end{figure}

There are 2 MHV vertices with quark-anti-quarks, one for a single
pair of quark-antiquark and one for two quark-anti-quark pairs
which are  shown in Fig.~\ref{mhvall}. There is no MHV vertex for
3 or more pair  of quark-antiquark. All these (non-MHV) amplitudes
should be computed from the above MHV vertices by drawing all
possible (connected) diagrams with only MHV vertices.

The explicit formulas for these MHV vertices (or amplitudes) are
given as follows:
\begin{eqnarray}
& & V(\Lambda_q^+, g_1^+, \cdots, g_I^-, \cdots, g_n^+,
\Lambda_{\bar q}^-) = -{\langle q, I\rangle \langle \bar q, I
\rangle^3 \over \langle q, 1\rangle \langle 1,2 \rangle \cdots
\langle n ,\bar q\rangle
\langle \bar q ,q  \rangle} , \\
& & V(\Lambda_q^-, g_1^+, \cdots, g_I^-, \cdots, g_n^+,
\Lambda_{\bar q}^+) = {\langle q, I\rangle^3 \langle \bar q, I
\rangle \over \langle q, 1\rangle \langle 1,2 \rangle \cdots
\langle n ,\bar q\rangle \langle \bar q ,q  \rangle} ,
\end{eqnarray}
for the single pair of quark-antiquark,  and for 2 quark-antiquark
pairs:
\begin{eqnarray}
 & &\hskip -1cm V(\Lambda_{q_1}^{h_1}, g_1, \cdots, g_{n_1},
 \Lambda_{\bar{ q}_2}^{-h_2}, \Lambda_{q_2}^{h_2}, g_{n_1+1},
 \cdots, g_{n}, \Lambda_{\bar{ q}_1}^{-h_1})  \nonumber\\
 &&= A_0(h_{ 1},h_{ 2})
 {\langle q_1, \bar{q}_2\rangle\over
\langle q_1, 1\rangle \langle  1,  2\rangle \cdots
 \langle n_1, \bar{q_2}\rangle}   \times
{\langle q_2, \bar{q}_1\rangle\over \langle q_2, n_1+1\rangle
  \cdots
 \langle n,\bar{q}_1\rangle}  ,
 \label{eqmhv4f}
\end{eqnarray}
where $A_0(h_{ 1}, h_{ 2})$ is given as follows:
\begin{eqnarray}
A_0(+,+)={\langle \bar{q}_1, \bar{q}_2 \rangle^2 \over \langle
q_1, \bar{q}_1 \rangle \langle q_2, \bar{q}_2\rangle}  , &&
A_0(+,-)=-{\langle \bar{q}_1, q_2 \rangle^2 \over \langle q_1,
\bar{q}_1 \rangle \langle q_2, \bar{q}_2\rangle} , \\
A_0(-,+)=-{\langle q_1, \bar{q}_2 \rangle^2 \over \langle q_1,
\bar{q}_1 \rangle \langle q_2, \bar{q}_2\rangle},  &&
A_0(-,-)={\langle q_1, q_2 \rangle^2 \over \langle q_1, \bar{q}_1
\rangle \langle q_2, \bar{q}_2\rangle} .
\end{eqnarray}
All these MHV amplitudes are given in terms of the ``holomorphic''
spinors of the external (on-shell) momenta. So we can use the same
off shell continuation given in \cite{Wittena} which we recalled
above. By including these fermionic MHV vertices, we can extend
the CSW rule of perturbative gauge theory  to gauge theories with
quarks and anti-quarks.   The propagator for both gluon and
fermion (quark or anti-quark) internal lines is just $1/p^2$, as
explained in \cite{GeorgiouKhoze}.  Helicity is conserved along a
fermion line. Because we assume that all quarks have different
flavor, the flowing of arrows must follow the directions given
exactly in Fig.~\ref{mhvall}. We use the rules given in
\cite{GeorgiouKhoze} to sew vertices by propagators.

Assume that in an MHV diagram, there are $n_i$ purely gluonic MHV
vertices with $i$ lines, $n_i^{2f}$ single pair of quark-antiquark
MHV vertices with $(i+2)$ lines (counting also the 2 fermion
lines, so actually only $i$ gluon lines), and $n_i^{4f}$ 2  pairs
of quark-antiquark MHV vertices with $(i+4)$ lines (counting also
the 4 fermion lines, so actually only $i$ gluon lines). For a
connected diagram with $m$ quark-antiquark pairs, the number of
positive helicity gluon $n_+$ and the number of negative helicity
gluon $n_-$ are given as follows \cite{WuZhub}:
\begin{eqnarray}
n_- & = & \sum_{i\ge 3} n_i + \sum_{i\ge 1}n_i^{2f} + \sum_{i\ge
0}
n_i^{4f} - (m-1),  \label{numbermm} \\
n_+ & = & \sum_{i\ge 3} (i-3) \, n_i + \sum_{i\ge 1}(i-1)\,
n_i^{2f} + \sum_{i\ge 0} (i+1)\, n_i^{4f} - (m-1) .
\label{numberpp}
\end{eqnarray}

\begin{figure}[ht]
    \epsfxsize=100mm%
    \hfill\epsfbox{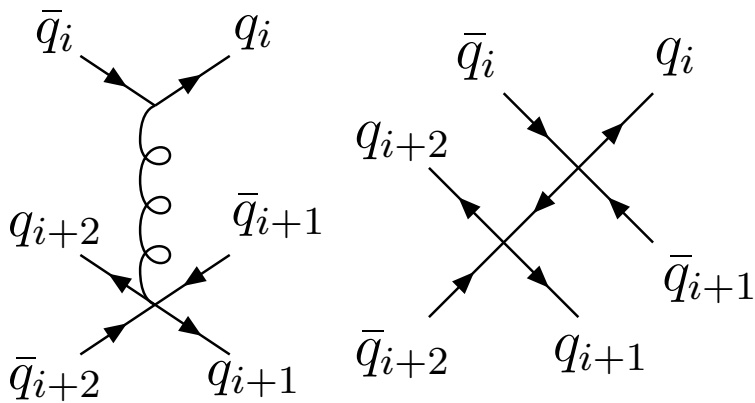}\hfill~\\
    \caption{The 2 different kinds of CSW diagrams  contributing to
    the purely fermionic amplitude with 3 quark-antiquark pairs. In these diagrams  $i$ can take $1,2,3$
    and the indices are understood as mod $3$.
    So totally there are $6$ diagrams.}
           \label{6quark}
   \end{figure}

These relations are particularly useful for analyzing diagrams
with fewer number of external gluons with positive helicity. For
the purely fermionic amplitudes with $3$ quark-antiquark pairs,
there are  only $2$ different kinds of diagrams as shown in
Fig.~\ref{6quark} \cite{WuZhub}. By using the extended CSW rules,
the partial amplitude can be written down very simply. We show in
the next section that this gives the right result for the
amplitude.

\section{The purely fermionic amplitudes with three quark-antiquark pairs}
As mentioned in the previous section,  we will compute the purely
fermionic amplitudes with three distinct (massless)
quark-anti-quark pairs.
 The color factor
is $\delta_{i_1 \bar i_2} \delta_{i_2 \bar i_3}\delta_{i_3 \bar
i_1}$, for a particular ordering of the quark-antiquarks. The
partial amplitude is denoted as
$A_6(\Lambda_{q_1}^{h_1},\Lambda_{\bar
q_2}^{-h_2},\Lambda_{q_2}^{h_2},\Lambda_{\bar
q_3}^{-h_3},\Lambda_{q_3}^{h_3},\Lambda_{\bar q_1}^{-h_1})$.

In this section, we will compute these partial amplitudes first by
using the CSW rules, then by using the Feynman rules. we will show
that these two results coincite up to a phase factor because of
the phase convention we used for the vertices and the propagators
in the CSW approach.

As mentioned above, in the CSW approach to compute this partial
amplitudes, there are only $2$ different kinds of CSW diagrams as
shown in Fig.~\ref{6quark}. (We note that these CSW diagrams
corresponding to the amplitudes with different helicity
configurations  are the same. This is different from many gluonic
amplitudes, where the CSW diagrams corresponding to the amplitudes
with different helicity configurations are different.)

There are $8$ kinds of helicity configurations. We can find some
relations among these amplitudes with different helicity
configurations.

This first relation is:
\begin{equation}
A_6(\Lambda_{q_1}^{h_1},\Lambda_{\bar
q_2}^{-h_2},\Lambda_{q_2}^{h_2},\Lambda_{\bar
q_3}^{-h_3},\Lambda_{q_3}^{h_3},\Lambda_{\bar
q_1}^{-h_1})=A_6(\Lambda_{q_2}^{h_2},\Lambda_{\bar
q_3}^{-h_3},\Lambda_{q_3}^{h_3},\Lambda_{\bar
q_1}^{-h_1},\Lambda_{q_1}^{h_1},\Lambda_{\bar q_2}^{-h_2}).
\end{equation}
This means that the amplitudes are invariant under the cyclic
permutation of the quark-anti-quark pairs.

The second one is the relation between the two amplitudes related
by charge conjugation:
\begin{equation}
A_6(\Lambda_{q_1}^{h_1},\Lambda_{\bar
q_2}^{-h_2},\Lambda_{q_2}^{h_2},\Lambda_{\bar
q_3}^{-h_3},\Lambda_{q_3}^{h_3},\Lambda_{\bar
q_1}^{-h_1})=-A_6(\Lambda_{q^{\prime}_3}^{-h_3},\Lambda_{ \bar
q^{\prime}_2}^{h_2},\Lambda_{q^{\prime}_2}^{-h_2},\Lambda_{\bar
q^{\prime}_1}^{h_1}, \Lambda_{q^{\prime}_1}^{-h_1},\Lambda_{\bar
q^{\prime}_3}^{h_3}),
\end{equation}
where $\lambda_{q^{\prime}_i}=\lambda_{\bar q_i},\lambda_{\bar
q^{\prime}_i}=\lambda_{q_i},\tilde\lambda_{q^{\prime}_i}=\tilde\lambda_{\bar
q_i}$,$\tilde\lambda_{\bar q^{\prime}_i}=\tilde\lambda_{q_i}$.

These relations can be easily verified case by case, either using
the CSW rules or using the Feynman rules.

So we only need to consider the case when $h_1=h_2=h_3=-1/2$ and
the case when $h_1=h_2=-1/2, h_3=1/2$. Other cases can be obtained
from these cases by permutation of the quark-anti-quark pairs
and/or charge conjugation transformation.

When $h_1=h_2=h_3=-1/2$, the amplitude is:
\begin{equation}
A^{CSW}_6(\Lambda_{q_1}^-,\Lambda_{\bar
q_2}^+,\Lambda_{q_2}^-,\Lambda_{\bar
q_3}^+,\Lambda_{q_3}^-,\Lambda_{\bar q_1}^+)=\sum_{i=1}^3
A^i+\sum_{i=1}^3 \tilde A^i,
\end{equation}

where

\begin{eqnarray}
A^i&=&-{\langle q_i,(\bar q_i\,q_i)\rangle^3 \, \langle \bar
q_i,(\bar q_i\,q_i)\rangle \over \langle \bar q_i,
q_i\rangle\,\langle q_i,(\bar q_i\,q_i)\rangle\,\langle (\bar
q_i\,q_i), \bar q_i\rangle}\,{1 \over (p_{\bar
q_i}+p_{q_i})^2}\nonumber
\\&& \times
{\langle q_{i+1}, q_{i+2}\rangle^2\over\langle \bar
q_{i+1},q_{i+1}\rangle\, \langle \bar
q_{i+2},q_{i+2}\rangle}\,{\langle q_{i+2}, \bar
q_{i+1}\rangle\over \langle q_{i+2}, (\bar q_i \,  q_i) \rangle
\langle (\bar q_i \, q_i), \bar q_{i+1} \rangle }\nonumber \\
&=&{\langle q_i,(\bar q_i\,q_i)\rangle^2  \over \langle \bar q_i,
q_i\rangle}\,{1 \over (p_{\bar q_i}+p_{q_i})^2}\, {\langle
q_{i+1}, q_{i+2}\rangle^2\over\langle \bar
q_{i+1},q_{i+1}\rangle\, \langle \bar
q_{i+2},q_{i+2}\rangle}\nonumber
\\&& \times{\langle q_{i+2}, \bar
q_{i+1}\rangle\over \langle q_{i+2}, (\bar q_i \,  q_i) \rangle
\langle (\bar q_i \, q_i), \bar q_{i+1} \rangle },
\end{eqnarray}

and
\begin{eqnarray}
\tilde A^i&=&-{\langle q_i, (\bar q_i \, \bar q_{i+1})\rangle^2
\over \langle \bar q_i, q_i\rangle \, \langle \bar q_{i+1}, (\bar
q_i\, q_i) \rangle} \,{1\over (p_{\bar q_i}+p_{q_i}+p_{\bar
q_{i+1}})^2}\\ \nonumber &&\times  {\langle q_{i+1},
q_{i+2}\rangle^2 \over \langle (\bar q_{i+2}\, q_{i+2}),
q_{i+1}\rangle \,\langle \bar q_{i+2}, q_{i+2}\rangle},
\end{eqnarray}
Here the expression $\langle(ij), k\rangle$ is defined as $\langle
\lambda_{p_i+p_j}, \lambda_k \rangle$, and the indices are
understood mod $3$.

From the momentum conservation we get
$\lambda_{q_i}\phi_{q_i}+\lambda_{\bar q_i}\phi_{\bar
q_i}+\lambda_{p_{q_i}+{p_{\bar q_i}}}=0$, so $\langle q_i,(\bar
q_i\,q_i)\rangle=\phi_{\bar q_i}\langle \bar q_i, q_i \rangle$
then
\begin{equation}
{\langle q_i,(\bar q_i\,q_i)\rangle^2  \over \langle \bar q_i,
q_i\rangle}\,{1 \over (p_{\bar q_i}+p_{q_i})^2}={\phi_{\bar
q_i}^2\langle \bar q_i,q_i \rangle\over (p_{\bar
q_i}+p_{q_i})^2}={\phi_{\bar q_i}^2\over [\bar q_i,q_i]}.
\end{equation}
Using this result, we can write $A^i$ as
\begin{eqnarray}
A^i={\phi_{\bar q_i}^2\over [\bar q_i, q_i]}\, {\langle q_{i+1},
q_{i+2}\rangle^2\over\langle \bar q_{i+1},q_{i+1}\rangle\,\langle
\bar q_{i+2},q_{i+2}\rangle}\,{\langle q_{i+2}, \bar
q_{i+1}\rangle\over \langle q_{i+2}, (\bar q_i \,  q_i) \rangle
\langle (\bar q_i \, q_i), \bar q_{i+1} \rangle }.
\end{eqnarray}

Similarly, when $h_1=h_2=-1/2,h_3=1/2$, the result given by the
CSW rules is

\begin{equation}
A^{CSW}_6(\Lambda_{q_1}^-,\Lambda_{\bar
q_2}^+,\Lambda_{q_2}^-,\Lambda_{\bar
q_3}^-,\Lambda_{q_3}^+,\Lambda_{\bar q_1}^+)=\sum_{i=1}^3
A^i+\sum_{i=1}^3 \tilde A^i.
\end{equation}

Now

\begin{eqnarray}
A^1=-{\phi_{\bar q_1}^2\over [\bar q_1,q_1]}{\langle q_2, \bar q_3
\rangle^2\over \langle \bar q_2, q_2 \rangle \, \langle \bar q_3,
q_3 \rangle}{\langle q_3,\bar q_2\rangle\over\langle q_3, (\bar
q_1\,q_1) \rangle \,\langle (\bar q_1\,q_1),\bar q_2 \rangle},
\end{eqnarray}

\begin{eqnarray}
A^2=-{\phi_{\bar q_2}^2\over [\bar q_2,q_2]}{\langle q_1, \bar q_3
\rangle^2\over \langle \bar q_3, q_3 \rangle \, \langle \bar q_1,
q_1 \rangle }{\langle q_1,\bar q_3\rangle\over\langle q_1, (\bar
q_2\,q_2) \rangle \, \langle (\bar q_2\,q_2),\bar q_3 \rangle},
\end{eqnarray}

\begin{eqnarray}
A^3=-{\phi_{q_3}^2\over [\bar q_3,q_3]}{\langle q_1, q_2
\rangle^2\over \langle \bar q_1, q_1 \rangle \, \langle \bar q_2,
q_2 \rangle }{\langle q_2,\bar q_1\rangle\over\langle q_2, (\bar
q_3\,q_3) \rangle \, \langle (\bar q_3\,q_3),\bar q_1 \rangle},
\end{eqnarray}

and
\begin{eqnarray}
\tilde A^1&=&{\langle (q_3 \, \bar q_1), q_1\rangle^2\over\langle
\bar q_1, q_1 \rangle\, \langle (\bar q_1\, q_1),
q_3\rangle}\,{1\over (p_{\bar q_2}+p_{q_2}+p_{\bar q_3})^2}\\
\nonumber &&\times {\langle q_2, \bar q_3 \rangle^2 \over \langle
\bar q_2, q_2 \rangle\,\langle \bar q_3, (\bar q_2\,
q_2)\rangle},\end{eqnarray}

\begin{eqnarray}
\tilde A^2&=&{\langle (q_3 \, \bar q_1), \bar q_3 \rangle^2\over
\langle \bar q_3, q_3\rangle\, \langle \bar q_1, (\bar q_3\,
q_3)\rangle}\, {1\over (p_{\bar q_3}+p_{q_3}+p_{\bar q_1})^2}
\\
\nonumber  &&\times {\langle q_1, q_2 \rangle^2\over \langle \bar
q_2, q_2 \rangle\,\langle (\bar q_2\, q_2), q_1 \rangle} ,
\end{eqnarray}

\begin{eqnarray}
\tilde A^3&=&{\langle q_1, (\bar q_1\, \bar q_2)\rangle^2\over
\langle \bar q_1, q_1 \rangle\, \langle \bar q_2, (\bar q_1 \,
q_1) \rangle}\, {1\over (p_{\bar q_1}+p_{q_1}+p_{\bar
q_2})^2} \\
\nonumber &&\times {\langle q_2, \bar q_3 \rangle^2\over \langle
\bar q_3, q_3 \rangle\, \langle (\bar q_3\, q_3), q_2\rangle}.
\end{eqnarray}

 \begin{figure}[ht]
    \epsfxsize=100mm%
    \hfill\epsfbox{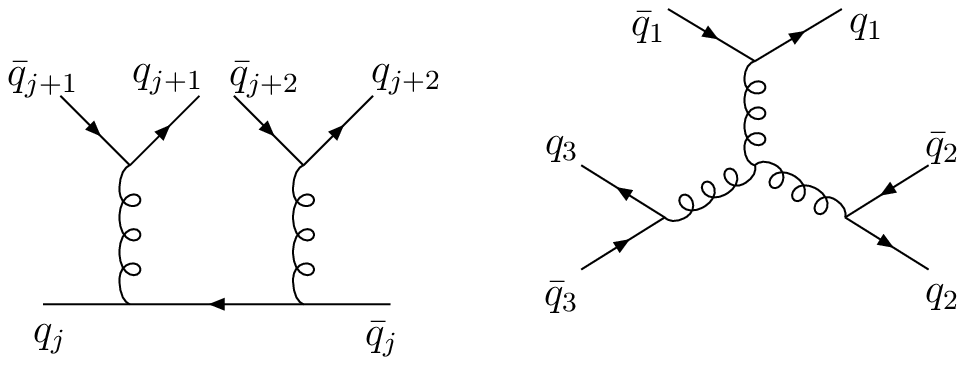}\hfill~\\
    \caption{The 2 different kinds of Feynman diagrams  contributing to
    the purely fermionic amplitude with 3 quark-antiquark pairs. In the left diagram  $j$ can take $1,2,3$
    and the indices are understood as mod $3$.
    So totally there are $4$ diagrams.}
     \label{feyn}
   \end{figure}

Now we will compute this amplitudes by using the Feynman rules
\footnote{These calculations have been done in \cite{Gunion}. We
thank Zvi Bern for reminding us of this.}. There are four Feynman
diagrams as shown in Fig.~\ref{feyn}. We will use the helicity
trick (see, for example, \cite{Dixon})\footnote{We note that the
convention for $[i,j]$ in \cite{Dixon} is different from the one
we used here by an extra $-$.}.

The result for $h_1=h_2=h_3=-1/2$ is
\begin{equation}
A^{Feynman}_6(\Lambda_{q_1}^-,\Lambda_{\bar
q_2}^+,\Lambda_{q_2}^-,\Lambda_{\bar
q_3}^+,\Lambda_{q_3}^-,\Lambda_{\bar q_1}^+)=\sum_{j=1}^3
A^j+\tilde A,
\end{equation}
where\footnote{The amplitudes we calculate are
$A_6(\Lambda_{q_1}^{h_1},\Lambda_{\bar
q_2}^{-h_2},\Lambda_{q_2}^{h_2},\Lambda_{\bar
q_3}^{-h_3},\Lambda_{q_3}^{h_3},\Lambda_{\bar q_1}^{-h_1})$
instead of $A_6(\Lambda_{\bar
q_1}^{-h_1},\Lambda_{q_1}^{h_1},\Lambda_{\bar
q_2}^{-h_2},\Lambda_{q_2}^{h_2},\Lambda_{\bar
q_3}^{-h_3},\Lambda_{q_3}^{h_3})$, this fact gives an extra $-$.}
\begin{eqnarray}
A^j&=&-\langle p_{q_j}^- |{i \over \sqrt 2}\gamma^{\mu} i
\left(p_{q_j}^{\rho}+p_{\bar q_{j+1}}^{\rho}+
p_{q_{j+1}}^{\rho}\right)\gamma_{\rho} {i \over \sqrt 2}\gamma^{\nu}|p_{\bar q_j}^-\rangle \nonumber\\
&&\times \langle p_{q_{j+1}}^-| {i \over \sqrt 2}\gamma_{\mu}
|p_{\bar q_{j+1}}^-\rangle \langle p_{q_{j+2}}^-| {i \over \sqrt
2}\gamma_{\nu} |p_{\bar q_{j+2}}^-\rangle {-i \over (p_{\bar
q_{j+1}}+p_{q_{j+1}})^2} {-i \over (p_{\bar
q_{j+2}}+p_{q_{j+2}})^2}\nonumber \\
&&\times{1\over (p_{q_j}+p_{\bar
q_{j+1}}+p_{q_{j+1}})^2}\nonumber\\
&=&i {\langle q_j, q_{j+1}\rangle\, [\bar q_j,\bar q_{j+2}]\over
(p_{\bar q_{j+1}}+p_{q_{j+1}})^2\,(p_{\bar
q_{j+2}}+p_{q_{j+2}})^2\,(p_{q_j}+p_{\bar
q_{j+1}}+p_{q_{j+1}})^2}\nonumber \\
&&\times \left(\langle q_j, q_{j+2}\rangle\,[q_j,\bar
q_{j+1}]+\langle q_{j+1}, q_{j+2}\rangle\,[q_{j+1},\bar
q_{j+1}]\right),\label{eq29}\end{eqnarray}

and

\begin{eqnarray}
\tilde A&=&- \langle p_{q_1}^-| {i \over \sqrt 2}\gamma_{\mu}
|p_{\bar q_1}^-\rangle
 \langle p_{q_2}^-| {i \over \sqrt 2}\gamma_{\nu} |p_{\bar q_2}^-\rangle
\langle p_{q_3}^-| {i \over \sqrt 2}\gamma_{\rho} |p_{\bar q_3}^-\rangle\nonumber\\
&&\times {i \over \sqrt 2}\left( 2 (k_{\bar q_1}+k_{q_1})_{\rho}
\eta_{\mu\nu}+
2 (k_{\bar q_2}+k_{q_2})_{\mu} \eta_{\nu\rho}+2 (k_{\bar q_3}+k_{q_3})_{\nu} \eta_{\rho\mu} \right)\nonumber\\
&&\times \prod_{j=1}^3 {-i \over  (p_{\bar q_j}+p_{q_j})^2} \nonumber \\
&=& {-i \over\prod_{j=1}^3 (p_{\bar q_j}+p_{q_j})^2}\sum_{k=1}^3
\langle q_k,q_{k+1}\rangle\,[\bar q_k,\bar
q_{k+1}]\nonumber \\
&&\times \left(\langle \bar q_k, q_{k+2}\rangle\,[\bar q_k, \bar
q_{k+2}]+\langle  q_k, q_{k+2}\rangle\,[ q_k, \bar
q_{k+2}]\right).\label{eq30}
\end{eqnarray}
The bras and kets in eqs.~(\ref{eq29}) and (\ref{eq30}) are
denoted by the momenta of corresponding particles.

When $h_1=h_2=-1/2,h_3=1/2$, we can similarly obtain the following
result by using the Feynman rules,

\begin{equation}
A^{Feynman}_6(\Lambda_{q_1}^-,\Lambda_{\bar
q_2}^+,\Lambda_{q_2}^-,\Lambda_{\bar
q_3}^-,\Lambda_{q_3}^+,\Lambda_{\bar q_1}^+)=\sum_{j=1}^3 A^j+
\tilde A.
\end{equation}

Now \begin{eqnarray}A^1&=&i{\langle q_1, q_2 \rangle\,[\bar q_1,
q_3]\over (p_{\bar q_2}+p_{q_2})^2\,(p_{\bar
q_3}+p_{q_3}^2)\,(p_{q_1}+p_{\bar q_2}+p_{q_2})^2}\nonumber\\
&& \times \left(\langle q_1,\bar q_3 \rangle\,[q_1,\bar
q_2]+\langle q_2,\bar q_3 \rangle\,[q_2,\bar q_2] \right),
\end{eqnarray}

 \begin{eqnarray}A^2&=&i{\langle q_2,\bar q_3 \rangle\,[\bar q_2,
\bar q_1]\over (p_{\bar q_3}+p_{q_3})^2\,(p_{\bar
q_1}+p_{q_1})^2\,(p_{q_2}+p_{\bar q_3}+p_{q_3})^2}\nonumber\\
&&\times \left(\langle q_2, q_1 \rangle\,[q_2, q_3]+\langle \bar
q_3, q_1 \rangle\,[\bar q_3, q_3] \right),
\end{eqnarray}

\begin{eqnarray}
A^3&=&i{\langle \bar q_3, q_2\rangle\, [q_3,\bar q_1]\over
(p_{\bar q_1}+p_{q_1})^2\,(p_{\bar
q_2}+p_{q_2})^2\,(p_{q_3}+p_{\bar q_1}+p_{q_1})^2}\nonumber\\
&&\times \left(\langle q_3, q_1\rangle\, [q_3,\bar q_2]+\langle
\bar q_1, q_1 \rangle\,[\bar q_1, \bar q_2] \right),
\end{eqnarray}

and

\begin{eqnarray}
\tilde A&=&-i \left(\langle q_1, q_2\rangle\, [\bar q_1, \bar
q_2]\left(\langle\bar q_3, \bar q_1 \rangle \,[q_3, \bar
q_1]+\langle\bar q_3, q_1 \rangle\, [q_3,
q_1]\right)\right. \nonumber\\
  & & \left. +\langle q_2, \bar q_3\rangle\, [\bar q_2, q_3]\left(\langle q_1,
\bar q_2 \rangle \,[\bar q_1, \bar q_2]+\langle q_1, q_2 \rangle\,
[\bar q_1, q_2]\right)\right. \nonumber\\
  & & \left. +\langle\bar q_3, q_1\rangle\,[q_3, \bar q_1]
\left(\langle q_2, q_3 \rangle\,[\bar q_2, q_3]+\langle q_2, \bar
q_3 \rangle\, [\bar q_2, \bar q_3] \right) \right)\nonumber\\
  && \times  {1\over \sum_{j=1}^3(p_{\bar
q_j}+p_{q_j})^2}.
\end{eqnarray}

There are the following constrains from the momentum conservation:
\begin{equation}
\sum_{i=1}^3\lambda_{q_i}^{\alpha}\tilde\lambda_{q_i}^{\dot
\alpha}+ \sum_{i=1}^3\lambda_{\bar
q_i}^{\alpha}\tilde\lambda_{\bar q_i}^{\dot \alpha}=0,\,
\alpha=1,2, \,\dot\alpha=1,2.
\end{equation}
From these constrains, we can solve $\tilde \lambda_{\bar
q_1}^{\dot \alpha}$ and $\tilde \lambda_{q_1}^{\dot \alpha}$ in
terms of other $\lambda$'s and $\tilde\lambda$'s. The result is
\begin{eqnarray}
\tilde \lambda_{\bar q_1}^{\dot
\alpha}&=&-\sum_{i=2}^3{\tilde\lambda_{q_i}^{\dot \alpha} \langle
\lambda_{q_i},\lambda_{q_1} \rangle+\tilde\lambda_{\bar q_i}^{\dot
\alpha} \langle \lambda_{\bar q_i},\lambda_{ q_1}
\rangle\over\langle\lambda_{\bar q_1},\lambda_{q_1} \rangle}, \nonumber \\
\tilde \lambda_{q_1}^{\dot
\alpha}&=&-\sum_{i=2}^3{\tilde\lambda_{q_i}^{\dot \alpha} \langle
\lambda_{q_i},\lambda_{\bar q_1} \rangle+\tilde\lambda_{\bar
q_i}^{\dot \alpha} \langle \lambda_{\bar q_i},\lambda_{\bar q_1}
\rangle\over\langle\lambda_{q_1},\lambda_{\bar q_1} \rangle}.
\end{eqnarray}
We noted that we don't treat $\tilde \lambda$ as the complex
conjugation of $\lambda$. So in fact, we have use analytic
continuation to the spacetime with signature $(2,2)$, after we
obtain our result we can go back the Minkowski space. By  using
these results and with the help of symbolic manipulation, we can
find that
\begin{equation}
A^{Feynman}_6=-i A^{CSW}_6,
\end{equation}
either in the case when $h_1=h_2=h_3=-1/2$ or in the case when
$h_1=h_2=-1/2,h_3=1/2$. From the argument above we know that we
can obtain the same result for all of the helicity configurations
as promised.

\vspace{5mm}
\noindent{\bf Acknowledgment}

We would like to thank Zvi Bern for useful discussion during
ichep'04 at Beijing and  Chuan-Jie Zhu for suggesting this topic
and discussion.

\end{document}